# The Exact Solution of Lagrange's Gyroscope

John Schutkeker

NLM Physics Corp.
Royal Oak, MI 48073

December 18, 2023

The closed form solution is found for the fully nonlinear dynamics of the gyroscope with a fixed point at the tip. The solution is found by using Cardano's formulae to factor a cubic, in the case where all roots are known to be real. From this, the nutation angle is solved first in terms of Jacobi's elliptic integral of the first kind. A simple change of variables then transforms the dependent variable of the remaining equations from time to the projection of the nutation angle onto the vertical axis. After this transformation, the remaining equations can be integrated exactly, giving solutions expressed in Jacobi's elliptic integrals of the third kind. Reduced energy, angular momentum, moment of inertia and Cardano's discriminant are defined. The thresholds are found, separating looping, cuspidial, and unidirectional domains of nutation.

# 1) Introduction

The first appearance of the gyroscope was as a child's toy in ancient times, the Jewish dreidle. Japanese children played with them, in their pre-industrial epoch, and they were popular with American children during the Great Depression. The first laboratory gyroscope was built in 1817 by Johann Bohnenberger, and it was named around 1852, by Foucault, who designed and built a precision device. Although Lagrange did not study it mathematically, it is still named after him, because of his development of a comprehensive mathematical theory of mechanics that describes everything.

Avionic applications include the autopilot, gyrocompass, artificial horizon and pre-GPS missile guidance. Since they're not subject to parasitic dynamics like precession, electrical gyros have replaced mechanical gyros, yet the mathematics of the mechanical problem remained unsolved.

Ginsburg [1] observed that the top can be solved with elliptic integrals, and MacMillan [2] implemented the beginnings of the closed form solution, with a graphical discussion of the necessity of solving the cubic equation. Sussman and Wisdom [6] observed that, because they are of first order only, the equations are solvable in closed form, and Goldstein [7] has observed that the solutions will be elliptic integrals.

This paper uses Cardano's formulae to factor Ginsberg's cubic equation and uses that result to integrate the equation of motion and find the closed form dynamics, in terms of elliptic integrals of the first and third kinds [3].

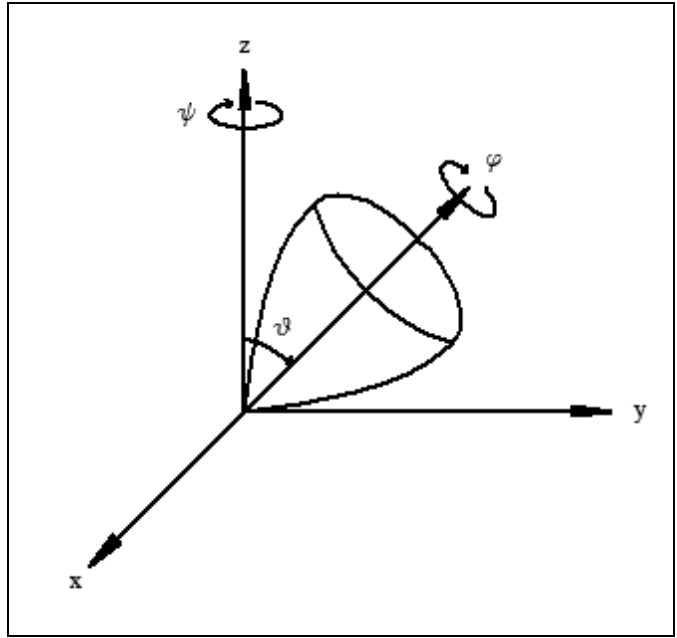
Figure 1 – Geometry of the gyroscope

**2) Dynamics**

Euler Angles

As shown in Fig. 1, the orientation in space of any solid body, including the spinning top, is described by three Euler angles. For a top, these angles are

$\theta$ - Tilt Angle (Pitch)
$\varphi$ - Spin Angle (Roll)
$\psi$ - Precession Angle (Yaw)

The large differences from the analogy to the plane are that the gravity vector is counter-clockwise by 90 degrees, and the roll rate would correspond to an impossibly high rate of spin.

ODE for Tile Angle

The starting point of this derivation is Ginsberg's ODE [1] for the time evolution of the tilt angle, $\theta$. This analysis picks up where Ginsberg's ended, with redefinition of his variables into more completely reduced parameters. Ginsberg's equation is

$$\dot{u}^2 = (\varepsilon - \lambda u)(1 - u^2) - (\beta_\psi - \beta_\varphi u^2), \qquad (1)$$

where $\omega_{\varphi o}$ and $\omega_{\psi o}$ are the initial values of angular speed in the $\varphi$ and $\psi$ directions, with units of angular speed [3], and the tilt angle is projected onto the vertical axis by making the change of variables $u = \cos\theta$.

The various parameters in this equation are defined as follows: the reduced energy, $\varepsilon$, is given by

$$\varepsilon = 2\frac{E}{I_\perp} - \frac{I_\perp}{I_\parallel}\beta_\varphi^2, \tag{2}$$

and the reduced angular momentum, $\lambda$, is defined as

$$\lambda = \frac{2mgL}{I_\perp}, \tag{3}$$

Later in this paper, these parameters will be reduced second time, meaning that we refer to this step as the first reduction step, and that these parameters are said to be "singly reduced." Of course, $I_\parallel$ and $I_\perp$ are the respective moments of inertia parallel and perpendicular to the spin axis, and $E$ and $L$ are the total energy and angular momentum, respectively, as usual in mechanics.

Expanding Ginsberg's equation into a cubic gives

$$\dot{u}^2 = u^3 - u^2\left(\frac{\varepsilon - \beta_\varphi^2}{\lambda}\right) - u\left(1 - \frac{\beta_\varphi\beta_\psi}{\lambda}\right) + \left(\frac{\varepsilon - \beta_\psi^2}{\lambda}\right). \tag{4}$$

It is shown how to factor the cubic in §3, which we skip for the moment, to find the actual solution.

Tilt Angle Solution

Ginsberg has observed that it is possible to isolate the independent and dependent variables ($u$ and $t$, respectively). Doing that and factoring the cubic gives

$$dt = \frac{du}{\sqrt{(u - u_1)(u - u_2)(u - u_3)}} \tag{5}$$

with $u_1$, $u_2$, and $u_3$ ranked in decreasing order of magnitude, ie. $u_1 > u_2 > u_3$.

The ODE has been reduced to an integral, and integrated to give

$$t = \int \frac{du}{\sqrt{(u-u_1)(u-u_2)(u-u_3)}} = \frac{2}{\sqrt{u_1 - u_3}} F(\alpha, p), \qquad (6)$$

with $F(\alpha,p)$ is Jacobi's elliptic integral of the first kind [2], defined as

$$F(\alpha, p) \equiv \int_0^\alpha \frac{d\alpha}{\sqrt{1 - p^2 \sin^2 \alpha}}, \qquad (7)$$

with parameters

$$\alpha = \sin^{-1} \sqrt{\frac{u_1 - u_3}{u_1 - u}}, \qquad (8)$$

and

$$p = \sqrt{\frac{u_1 - u_2}{u_1 - u_3}}. \qquad (9)$$

Change of Variables

Ginsberg's ODE's [1] for the precession and spin angles are

$$\frac{d\psi}{dt} = \frac{\beta_\psi - \beta_\varphi \cos\theta}{\sin^2\theta}, \qquad (10)$$

$$\frac{d\varphi}{dt} = \frac{\beta_\varphi(\hat{I}\sin^2\theta + \cos^2\theta) - \beta_\psi \cos\theta}{\sin^2\theta}, \tag{11}$$

where the reduced moment of inertia, $\hat{I} \equiv I_\perp/I_\parallel$, has been defined.

These last two equations are simplified by using the chain rule to apply the change of variables, $u = \cos\theta$,

$$\frac{d\varphi}{du} = \frac{d\varphi/dt}{du/dt}, \tag{12}$$

where $du/dt$ is obtained from the original ODE for $\theta$,

$$\frac{du}{dt} = \sqrt{(u-u_1)(u-u_2)(u-u_3)}. \tag{13}$$

The transformed ODE's for $\varphi$ and $\psi$ are then

$$\frac{d\psi}{du} = \frac{\beta_\psi - \beta_\varphi u}{(1-u^2)\sqrt{(u-u_1)(u-u_2)(u-u_3)}}, \tag{14}$$

$$\frac{d\varphi}{du} = \frac{1}{\sqrt{(u-u_1)(u-u_2)(u-u_3)}}\left(\hat{I}\beta_\varphi - \frac{u(\beta_\psi - \beta_\varphi u)}{(1-u^2)}\right). \tag{15}$$

Precession Angle

Directly integrating the ODE for the precession angle gives

$$\psi(u) = \frac{1}{\sqrt{u_3 - u_1}}\left(\frac{\Pi_+}{u_1 + 1} + \frac{\Pi_-}{u_1 - 1}\right), \tag{16}$$

where

$$\Pi_{\pm} \equiv \Pi\left(\frac{u \pm 1}{u_3 - u_1}, \lambda, \frac{1}{p^2}\right), \tag{17}$$

and $\Pi(n, x, m)$ is Jacobi's elliptic integral of the third kind [3], defined as

$$\Pi(n, x, m) \equiv \int_1^x \frac{dx}{(1 - n\sin^2(x))\sqrt{1 - m\sin^2(x)}}, \tag{18}$$

with

$$\lambda \equiv \sin^{-1}\sqrt{\frac{(u_1 - u_2)}{(u_1 - u)}}. \tag{19}$$

Spin Angle

Directly integrating the ODE for the spin angle gives

$$\varphi(u) = \frac{2\hat{I}\beta_\varphi}{\sqrt{u_1 - u_3}} F + \frac{2u_1(\beta_\psi - u_1\beta_\varphi)F + \sum_\pm (u_1 \pm 1)(\beta_\psi \pm \beta_\varphi)\Pi_\mp}{(u_1^2 - 1)\sqrt{u_2 - u_1}}, \tag{20}$$

and from Eq. 6, we see that the first term is just the familiar advance in rotational angle from the spin of the top.

$$\frac{2\hat{I}\beta_\varphi}{\sqrt{u_1 - u_3}} F = \hat{I}\beta_\varphi t. \tag{21}$$

### 3) Cubic Equation

To prepare Ginsberg's equation for factoring, write it as

$$a_\psi - a_{\psi\varphi} u - a_\varphi u^2 + u^3 = 0, \qquad (22)$$

where $a_\varphi \equiv (\varepsilon - \beta_\varphi^2)/\lambda$, $a_\psi \equiv (\varepsilon - \beta_\psi^2)/\lambda$, $a_{\varphi\psi} \equiv 1 - \beta_\psi \beta_\varphi/\lambda$, and the order of the subscripts on $a_{\varphi\psi}$ is unimportant. These parameters are all energies, and energy is normalized according to $a = \varepsilon/\lambda$.

This reduces the number of control variables from four ($\varepsilon, \lambda, \beta_\varphi, \beta_\psi$) to three ($a_\varphi, a_\psi, a_{\varphi\psi}$), by eliminating angular momentum.

#### Cardano's Formulae

Cardano's Formulae [4] define the following terms,

$$Q = -\frac{a_{\varphi\psi}}{3} + \left(\frac{a_\varphi}{3}\right)^2, \qquad (23)$$

and

$$R = -\frac{a_\psi}{2} - \frac{a_{\varphi\psi} a_\varphi}{2 \cdot 3} - \left(\frac{a_\varphi}{3}\right)^3.$$

$D'$ is Cardano's discriminant, given by

$$D \equiv Q^3 + R^2. \qquad (25)$$

#### Real Roots

Since $u = \cos\theta$ represents an actual angle of tilt, complex solutions are unphysical and are simply discarded. This occurs when $D < 0$, causing the factoring to become uniquely simple. The cubic must always have two real critical points, and if it doesn't, it means one of two things.

  1) The top has two critical points, but they're both complex. That means that it is spinning too slowly to allow it to stand up.

  2) The top has no critical points, because they have merged in an $A_2$ catastrophe [5]. That means that the top is unconditionally unstable, at all values of spin/energy, no matter how large. It can never balance under any circumstance, because it is much too tall and narrow. The short, wide top is extremely stable, whereas the tall thin top is extremely unstable.

For real roots, Cardano gives

$$\Theta \equiv Cos^{-1}\left(\frac{R}{(-Q)^{3/2}}\right), \qquad (26)$$

and if we define the ordered sequence, $m \equiv \{-1, 0, 1\}$, we can write the three roots as

$$u_m = -\frac{a_\varphi}{3} + 2\sqrt{-Q}\cos\left(\frac{\Theta}{3} + \delta_m\right), \qquad (27)$$

where $\delta_m \equiv 2\pi m/3.$

The formula for the cosine of a sum gives

$$\cos(\Theta + \delta_m) = \frac{|m|}{2}\cos\left(\frac{\Theta}{3}\right) - m\frac{\sqrt{3}}{2}\sin\left(\frac{\Theta}{3}\right). \qquad (28)$$

Unfortunately, because of the $\Theta/3$ term in the argument of the cosine, this cannot be reduced further, by using $\cos(\cos^{-1}(\Theta)) = 1$ to eliminate the trigonometric functions, so the calculation of the roots ends at this point.

If there were a trigonometric identity, analogous to the half-angle formulae, for one-third of an angle, it could be used here, but apparently no such identity exists. It has been stated that this is because of the impossibility of trisecting an angle with compass and straight edge, but this is probably not true, since modern mathematics provides us with a lot more tools than just a compass and a straight-edge.

It is worth noting that the discriminant in Cardano's Law can be reduced one step further, to

$$D' \equiv (1 + Q^3/R^2). \qquad (29)$$

Since all the sign information is contained completely within $Q^3$, the discriminant will be negative, and roots imaginary when

$$Q < -R^{2/3}. \qquad (30)$$

This is the same as finding the inflection point of Ginsberg's Cubic, which occurs at $a_\varphi/3$. Substituting the definitions of $Q$ and $R$ into this gives

$$-\frac{a_{\varphi\psi}}{3} + \left(\frac{a_\varphi}{3}\right)^2 = -\left[-\frac{a_\psi}{2} - \frac{a_{\varphi\psi}}{2}\frac{a_\varphi}{3} - \left(\frac{a_\varphi}{3}\right)^3\right]^{2/3}, \qquad (32)$$

which is the same as saying $a_\varphi/3 = 0$. This might be verified by taking the cube of both sides, then reducing the result with simple algebra.

Sorting the Roots

The roots must be sorted, to map $u_m$ from §3 onto $\{u_1, u_2, u_3\}$ from §2, but it is unknown how to do this, leaving the task beyond the scope this paper. The question of sorting the roots is, in fact, a second mathematical problem that must be researched and solved, as the topic of a future paper.

Domains of Precession

Ginsberg [1] has shown that $u_3 > 1$, which is also unphysical, because $u$ is a cosine. This means that, although $u_3$ appears as a term in the elliptic integrals, domains of precession are governed only by $u_1$ and $u_2$. Ginsberg also defines $u^* \equiv \beta_\psi/\beta_\varphi$, to define the domains. The domains are

| | |
|---|---|
| Unidirectional | $(u_1 \text{ or } u_2) < u^*$ |
| Looping | $u_1 < u^* < u_2$ |
| Cuspidial | $u^* = u_2$ |

## 4) Summary

We have derived the closed for solution to the dynamical equations of the spinning top, written in terms of the three Euler angles. This procedure is as follows,

1) Factor Ginsberg's cubic equation with Cardano's formulae,
2) Solve the nutation angle in terms of Jacobi's elliptic integral of the first kind, $F(t)$.
3) Reduce the number of control parameters was from four to three by defining the normalized energy, angular momentum, and moment of inertia.
4) Change the dependent variable from time to nutation angle in the differential equations for the precession and spin angles.
5) Solve those two parameters in terms of Jacobi's elliptic integrals of the third kind, $\Pi(n, x, m)$.
6) Write the domains of precession in terms of the new parameters.

## 5) Conclusion

This work has finally found an exact solution to a foundational gyro-dynamics problem which has languished unsolved for 210 years, since Lagrange first became aware of it near the end of his life, ca. 1813. Any new problems that can be built upon this are now open to investigation, analysis and the search for a solution. Wherever Lagrange's gyroscope is found, in any application of mechanical engineering, that system can now be analyzed and solved mathematically with an exact formula.

A concept for a more complex configuration, envisioned by the author, is that gyroscopes can theoretically be stacked on top of each other, the dynamics of which would be built up from mathematical building blocks based on this result. Such a system would exhibit eigenmodes and domains of stability, instability, and chaos. Analysis would be done by using linear algebra to build matrices of the dynamical equations of the individual systems. Assume solutions of the form discovered here, and see what equations result from that approach.

Of course, the starting point would be to try to solve the family of equations for only two gyroscopes, and then attempt to extend that solution to 3, then 4, and finally an arbitrary number of gyroscopes, in the manner typical of all mathematical problems in linear algebra and calculus, which are typically built up by stages.

One can also imagine that the axes of such a row of gyroscopes need not be oriented parallel to the gravity vector, but if oriented perpendicular to it, or at an arbitrary angle, the tips would have to be tightly fastened by rotating mechanical pivots, to them from coming apart.

The primary design problem would be incorporating motors and power supply, to keep the gyros running continuously. This problem was solved during the cold war, on the huge gyroscopes that were once used as the angular position sensor on ballistic missiles of the 60's and 70's. Such machines were closely guarded military secrets, but hopefully, their designs are available now, since they have all been completely replaced by electronic gyros, with no parasitic mechanical behaviors, like the precession, nutation, and vibration of mechanical gyros.

The objective would be to build a "toy," to act as a demonstrator, for achieving proof-of-concept. For this it would be infinitely more sensible to use small gyroscopes, like the ones sold at science museum gift shops, perhaps even somewhat smaller. However, attaching drive motors and power supplies would drastically change the inertia tensor, dominating the dynamics, and making the construction of such a machine difficult and expensive. I will probably remain a *gedanken* experiment, forever.

The *gedanken* experiment could further be extended to attach the machine to a pair of robot hands of another, external machine approximating the movement of the hands (and arms) of a person. We can be thankful that it can never be constructed in real life, because giving it to a real person to hold would be a terrible idea, since injury would most likely result.

In an unstable domain, the expected possibility is that this snake-like machine might abruptly collapse inward into itself, tangling itself up into a knot-like structure that would have a complex shape, but uninteresting dynamics, being effectively a machine that was jammed up.

But in domains where this does not happen, it would be expected that the chaotic dynamics of this serpent-like construct would be complex, ever changing, never repeating and extremely beautiful. And finally, in domains of low tilt angle, it would be expected that the dynamics would be simple, familiar, uncomplicated eigenmodes with a wavelike structure that is stable and non-repeating.

This analysis will not be attempted here, nor at any time in the future, so these predictions are left as challenges to computer programmers, to try to simulate, for the purpose of trying to discover a new fractal. Hopefully, the readers of this work will find these ideas to be stimulating and challenging, and will make the attempt to either do these mathematical analyses, write the computer simulations, or both, to try to unify theory and numerical experiment.

It would surely be a fascinating and engaging exercise, probably being extremely interesting to the general public, and certainly to professionals in science, math and computer programming. Other than being a toy, it seems doubtful that anybody could ever find a way to use either the device or a simulation program to make money, but human ingenuity at making money is boundless. At the very least it would be an amusing computer animation, fun to watch on YouTube or TikTok, and probably earn small money for the website. It would surely be an amusing segment on the TV news, or a science TV show, like Nova.

## Bibliography

[1] Ginsberg, J. H., *Advanced Engineering Dynamics*, Cambridge University Press, 2$^{nd}$ ed., 1998.

[2] MacMillan, W. D., *Dynamics of Rigid Bodies*, Dover Press, 1936.

[3] Bradbury, T. C., *Theoretical Mechanics*, Wiley, 1968.

[4] Press, *et al*, *Numerical Recipes*, Cambridge University Press, 1$^{st}$ ed., 1986.

[5] Gilmore, R., *Catastrophe Theory for Scientists and Engineers*, Dover Press, 1981.

[6] Sussman G. J. and Wisdom, J., *Structure and Interpretation of Classical Mechanics*, MIT Press, 1$^{st}$ ed., 2001.

[7] Goldstein, H. *Classical Mechanics,* 2$^{nd}$ ed., Addison-Wesley, 1980.